\documentclass{article}

\usepackage{arxiv}

\usepackage[utf8]{inputenc} 
\usepackage[T1]{fontenc}    
\usepackage{hyperref}       
\usepackage{url}            
\usepackage{booktabs}       
\usepackage{amsfonts}       
\usepackage{nicefrac}       
\usepackage{microtype}      
\usepackage{graphicx}
\usepackage{amssymb}
\usepackage{amsmath}
\usepackage{lineno}
\usepackage{hyperref}
\usepackage{listings}
\usepackage{subfigure}
\usepackage{comment}
\usepackage{enumitem}
\usepackage{multirow}
\usepackage[dvipsnames]{xcolor}
\usepackage{booktabs} 
\usepackage{siunitx} 
\usepackage{soul} 

\DeclareMathOperator*{\argmin}{arg\,min}
\newcommand{\g}[1]{{\color{gray}{#1}}}

\title{IntraSeismic: a coordinate-based learning approach to seismic inversion}

\author{
  Juan Romero \\
  KAUST\\
  Thuwal, Kingdom of Saudi Arabia \\
  \texttt{juan.romeromurcia@kaust.edu.sa}\\
   \And
  Wolfgang Heidrich \\
  KAUST\\
  Thuwal, Kingdom of Saudi Arabia \\
  \texttt{wolfgang.heidrich@kaust.edu.sa}\\
   \And
  Nick Luiken \\
  KAUST\\
  Thuwal, Kingdom of Saudi Arabia \\
  \texttt{nicolaas.luiken@kaust.edu.sa}\\
  \And
  Matteo Ravasi \\
  KAUST\\
  Thuwal, Kingdom of Saudi Arabia \\
  \texttt{matteo.ravasi@kaust.edu.sa}\\
  }
  
\begin{document}

\chead{IntraSeismic: a coordinate-based learning approach to seismic inversion}

\maketitle

\begin{abstract}
Seismic imaging is the numerical process of creating a volumetric representation of the subsurface geological structures from elastic waves recorded at the surface of the Earth. As such, it is widely utilized in the energy and construction sectors for applications ranging from oil and gas prospection, to geothermal production and carbon capture and storage monitoring, to geotechnical assessment of infrastructures. Extracting quantitative information from seismic recordings, such as an acoustic impedance model, is however a highly ill-posed inverse problem, due to the band-limited and noisy nature of the data. This paper introduces IntraSeismic, a novel hybrid seismic inversion method that seamlessly combines coordinate-based learning with the physics of the post-stack modeling operator. Key features of IntraSeismic are i) unparalleled performance in 2D and 3D post-stack seismic inversion, ii) rapid convergence rates, iii) ability to seamlessly include hard constraints (i.e., well data) and perform uncertainty quantification, and iv) potential data compression and fast randomized access to portions of the inverted model. Synthetic and field data applications of IntraSeismic are presented to validate the effectiveness of the proposed method. 
\end{abstract}

\begin{figure*}[ht!]
    \includegraphics[width=\textwidth]{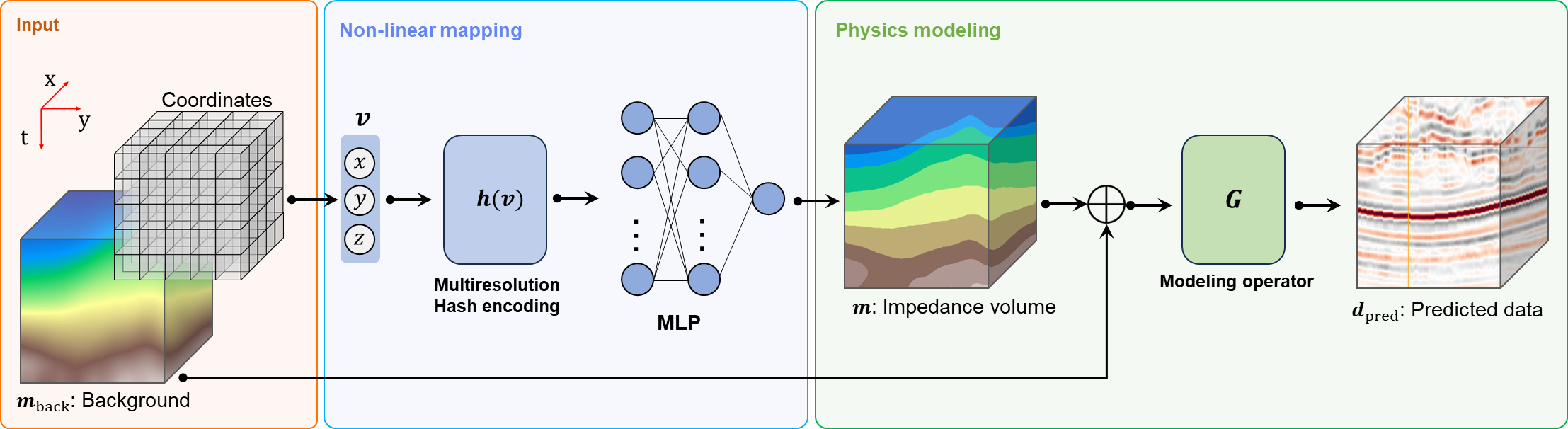}
    \caption{Schematic representation of the IntraSeismic framework: The input to IntraSeismic consists of spatial coordinates that are processed through a nonlinear mapping module. Initially, this module maps the input into a higher-dimensional space of trainable feature vectors using multiresolution hash encoding. Subsequently, these vectors are input to a MLP that outputs the impedance model. In the subsequent physics modeling module, the output from the MLP is added to the background model. A modeling operator then computes the predicted data from the impedance model, which is then incorporated into the loss function.}
    \label{fig:arch}
\end{figure*}

\section{Introduction}

\label{sec:intro}

Reflection seismology is the dominant geophysical method used to study the subsurface \cite{kearey_2002}, making it an indispensable tool for industries seeking to produce detailed insights into the Earth's interior. Seismic imaging, similar to non-light-of-sight imaging \cite{faccio2020}, relies on the principle of emitting waves from a source and interpreting the back-scattered signals to gain information about objects that are not directly visible. A variety of seismic imaging techniques have been developed over the years and found applications in several industrial sectors to aid oil and gas prospecting \cite{telford_applied_1990, sheriff_exploration_1995, Yilmaz}, geotechnical assessment in civil engineering \cite{oz_engineering_2015}, construction projects \cite{steeples_seismic_1988}, and more recently, geothermal production and carbon capture and storage monitoring~\cite{lumley_4d_2010}. Seismic inversion represents the foremost method for deriving quantitative information from seismic data \cite{tarantola_inversion_1984}. A seismic inverse problem endeavors to approximate the elastic parameters of the subsurface, such as wave propagation velocities and density, from either raw or processed seismic data. Even when this process is carried out post-migration (i.e., after the raw dataset has been imaged), the associated inverse problem, called \textit{post-stack seismic inversion}, is highly ill-posed due to the limited range of frequencies excited in a seismic experiment and the inevitable presence of noise in the data. 

\indent Historically, this problem has been tackled through three distinct families of methods: \textbf{1.} variational methods, \textbf{2.} deep learning-based approaches, and \textbf{ 3.} hybrid methods. Variational methods aim to minimize a data-misfit term accompanied by one or more regularization terms that guide the solution to a physically plausible outcome. A standard regularization choice is Tikhonov regularization~\cite{Tikhonov1977}, which favors the retrieval of a smooth subsurface model. More recently, total-variation regularization~\cite{rudin_nonlinear_1992} has been used to enforce a layered structure of the subsurface, which is a more realistic representation of the acoustic impedance~\cite{Gholami, Wang}. Deep learning-based approaches, on the other hand, leverage the prowess of neural networks to learn a direct mapping between the seismic data and acoustic impedance model, providing that a suitable training dataset is available. Various architectures have been used so far in the literature, including convolutional neural networks~\cite{biswas_prestack_2019,das_convolutional_2019}, temporal convolutional networks~\cite{smith_robust_2022}, and attention-based networks~\cite{wu_seismic_2022}. However, seismic applications deal with a shortage of labeled datasets because the ground truth is inaccessible except for specific locations along drilled wells (see Section \ref{methodology} for further details). Consequently, the application of supervised learning approaches is limited in this domain and is feasible only when a dense well coverage is available \cite{ravasi_deep_2023}. Hybrid seismic inversion methodologies harness the strengths of both paradigms by amalgamating the foundational geological knowledge and physics-based modeling with the pattern recognition capabilities of deep learning techniques. Examples of hybrid methods are Deep Image Prior (DIP)~\cite{ulyanov_deep_2020}, which uses a neural network as a nonlinear preconditioner in the solution of an inverse problem and has been successfully applied to the post-stack seismic inversion problem \cite{lipari_post-stack_2021}. Similarly, the Plug-and-Play (PnP) framework~\cite{venkatakrishnan_plug-and-play_2013} has recently been proposed to solve inverse problems using a denoiser to act as regularizer; this concept has led to the seamless inclusion of pre-trained deep denoisers~\cite{zhang_plug-and-play_2022} in the solution of inverse problems of any kind: \cite{romero_plug_2022} showed that this is also beneficial to 2D post-stack seismic inversion, achieving state-of-the-art results. For an extensive comparison of some of the previously described methods, we refer the reader to \cite{ravasi_deep_2023, ravasi_chapter_2023}. 

\indent These hybrid methods represent a forward-thinking direction for seismic inversion applications. Model-driven components ensure inversion results to be grounded in well-understood geophysical principles, providing interpretable and physically plausible solutions. Simultaneously, deep-learning-based components can capture complex, nonlinear relationships in the data and/or model that might be overlooked or overly simplified in purely model-driven techniques. In this spirit, we introduce IntraSeismic, a hybrid method that is able to:
\begin{itemize}
    \item Leverage the power of coordinate-based learning to solve ill-posed seismic inverse problems whilst retaining the physics of the seismic modeling operator through a modular framework.

    \item Seamless integrate hard constraints in the seismic inversion process. 

    \item Perform uncertainty quantification via Monte Carlo Dropout.

    \item Provide a compact representation of subsurface models, serving as an efficient data compression mechanism.
\end{itemize}
As shown by the numerical examples, IntraSeismic surpasses standard variational methods and other hybrid approaches in the inversion of both 2D and 3D post-stack seismic data in terms of reconstruction quality and convergence. This paper begins with an overview of post-stack seismic inversion, coordinate-based learning, and its application to inverse problems. Subsequently, we detail the components of the IntraSeismic framework and demonstrate the effectiveness of the proposed method on two synthetic datasets, namely the 2D Marmousi model \cite{brougois_marmousi_1990} and a subvolume of the 3D SEAM Phase 1 model \cite{Fehler2011}, and the Sleipner field dataset \cite{furre2017}. Finally, we illustrate the ability of IntraSeismic to include hard constraints, perform uncertainty quantification, and act as a flexible data compressor.

\section{Background and related work}

\paragraph{Post-stack seismic inversion} 
Post-migration seismic data (or seismic images) can be mathematically represented via the so-called convolutional model \cite{goupillaud_approach_1961}. This entails the convolution of a source function or wavelet $w(t)$ with the earth reflectivity series $r(t)$:
\begin{equation}
    d(t)=w(t) * r(t),
    \label{eq1}
\end{equation}
where $t$ indicates the time axis, $r(t)=(I_p(t+\Delta t) - I_p(t)) / (I_p(t+\Delta t) + I_p(t))$, and $I_p$ corresponds to the acoustic or $\mathrm{P}$-wave impedance. As shown in \cite{Stolt1985}, a logarithmic approximation holds for small to medium contrasts between geological layers and renders Equation~\ref{eq1} linear in $I_p(t)$:
\begin{equation}
    r(t)=\frac{1}{2} \frac{d}{d t} \log (I_p(t)).
\end{equation}
Hence, post-stack seismic data can be modeled as
\begin{equation}
    d(t)= \frac{1}{2} w(t) * \frac{d}{d t} \log (I_p(t)),
    \label{eqn:zero_offset}
\end{equation}
which can be written in compact matrix-vector notation:
\begin{equation}
    \mathbf{d} = \mathbf{W D m},
    \label{eqn:linear_poststack}
\end{equation}
where $\mathbf{W}$ represents a block-Toeplitz convolution matrix, which encapsulates the seismic wavelet, while $\mathbf{D}$ is the first-order derivative operator. Post-stack seismic inversion transforms seismic images $\mathbf{d}\in \mathbb{R}^{N_y N_x N_t}$ into quantitative estimates of the subsurface's acoustic impedance $\mathbf{m}\in \mathbb{R}^{N_y N_x N_t}$ \cite{Oldenburg1983, Russell1991}, and it is expressed as:

\begin{equation}
    \mathbf{m^*} = \argmin_{\mathbf{m}} \frac{1}{2}\left\|\mathbf{G m} - \mathbf{d}\right\|_2^2 + \mathcal{R}(\mathbf{m}),
    \label{eqn:poststack_inv}
\end{equation}
where $\mathbf{G} = \mathbf{W D}$ and $\mathcal{R}$ corresponds to the chosen regularization term.
\newline
\newline
\textbf{Coordinate-based learning for inverse problems}
Coordinate-based learning, wherein neural networks directly utilize spatial coordinates as input, has emerged as a focal point in recent deep learning advancements~\cite{tancik_fourier_2020}. This approach has achieved state-of-the-art results in various domains, including image and shape representation~\cite{stanley_compositional_2007, chen_learning_2019} and volume density and view synthesis~\cite{mildenhall_nerf_2020}, to name a few. Neural Radiance Fields (NeRFs)~\cite{mildenhall_nerf_2020} is an example of pioneering work in this domain, which has showcased the potent application of spatial coordinates in rendering photo-realistic 3D scenes.

\indent A recent notable contribution to the inverse problems field is the IntraTomo framework~\cite{zang_intratomo_2021}. IntraTomo synergizes a coordinate-based multi-layer perceptron (MLP) to estimate a density field with a subsequent geometry refinement module that solves an optimization problem to refine the details in such a density field. The data- and model-driven combination of IntraTomo has achieved state-of-the-art results in computer tomography. In the domain of seismic inversion, a parallel methodology has been suggested, involving the inversion of full-waveform seismic data using spatial coordinates as inputs to an MLP with variable activation functions \cite{sun2023}.

\indent In the present study, we draw inspiration from the IntraTomo and further develop this concept to produce a new technique called "IntraSeismic". In IntraSeimic, we take advantage of multi-resolution hash encoding with trainable feature vectors as a way to enrich the expressiveness of the network's input~\cite{muller_instant_2022}. As such, we are able to employ a very small, streamlined MLP architecture with two layers of 64 neurons each. Finally, we assimilate the 'geometry refinement module' of IntraTomo, by directly embedding model-based regularization terms in the neural network's loss function. Consequently, IntraSeismic emerges as a compact and computationally efficient framework that is tailor-made for solving highly ill-posed inverse problems in geophysics.

\section{The IntraSeismic framework}
    \label{methodology}

IntraSeismic integrates a self-supervised, coordinate-based learning module with a modeling operator to invert seismic data for acoustic impedance subsurface models (Figure \ref{fig:arch}). The proposed framework primarily uses coordinates as input, which first undergo a multi-resolution hash encoding process \cite{muller_instant_2022}. Next, the encoded coordinates are passed through an MLP, whose output is subsequently fed to the seismic modeling operator to allow evaluation of the loss function:
\begin{equation}
    \begin{aligned}
    \argmin_{\Theta} \frac{1}{2}\left\|\mathbf{G} \boldsymbol{m}_{\Theta} - \mathbf{d}\right\|_2^2 + \mathcal{R}(\boldsymbol{m}_{\Theta}) \\
    \boldsymbol{m}_{\Theta}=  \mathbf{F}_{\Theta}(\boldsymbol{x}) + \boldsymbol{m}_{\text{back}} \hspace{1cm},
    \label{eqn:intraseismic}
    \end{aligned}
\end{equation}
where \(\boldsymbol{m}_{\Theta}\) denotes the subsurface acoustic impedance volume obtained from the inversion process. \(\mathbf{F}_{\Theta}(\boldsymbol{x})\) is the network with \(\Theta\) representing the learnable parameters from both the hash table and the MLP network, and \(\mathbf{x}\in \mathbb{R}^{N_y, N_x, N_t \times 3}\) corresponds to the spatial coordinates of the volume of interest. These coordinates can be processed in IntraSeismic either as a single batch or in mini-batches. The term \(\boldsymbol{m}_{\text{back}}\) refers to the background model. The regularization term, \(\mathcal{R}(\boldsymbol{m}_{\Theta})\), employs anisotropic TV regularization on the model \(\left(\mathcal{R}(\boldsymbol{m}_{\Theta}) = \left\|\boldsymbol{m}_{\Theta}\right\|_{TV}\right)\) to achieve a high-resolution layered representation of the subsurface. Our experiments demonstrate improved performance when incorporating additional regularization on the network's output, such as sparsity-promoting regularization \(\mathcal{R}(\boldsymbol{m}_{\Theta}) = \left\|\boldsymbol{m}_{\Theta}\right\|_1\), or dynamically scheduled $L_2$ regularization \(\mathcal{R}(\boldsymbol{m}_{\Theta}) = \beta_k \left\|\boldsymbol{m}_{\Theta}\right\|_2^2\), where the regularization weight \(\beta\) diminishes across iterations \(k\) (for further details refer to the supplementary material). In the following text, we elucidate the different components of IntraSeismic.

\textbf{Input coordinates:} A post-stack seismic dataset is a 3D volume ($V$) of spatial size $N_y, N_x, N_t$ that can be represented via voxel discretization.  Each voxel $p_i$ within the volume $V$ is uniquely defined by its spatial coordinates $\boldsymbol{x_i} = (y, x, t)$. As shown in equation \ref{eqn:intraseismic}, the collection of all coordinates of the volume of interest represents the input of the IntraSeismic framework. Note the third dimension $N_t$ is time samples, as seismic data is recorded in time.

\textbf{Background Model:} A background or initial prior model plays a pivotal role in seismic inverse problems. This model must contain the low-frequency components of the subsurface impedance model, which are not captured in the data. In variational methods, the purpose of the background model is to steer the inverse process towards geologically consistent solutions. When using synthetic data, the background model is typically a smoothed version of the ground truth. In contrast, for field data, the background model is derived from an interpolated, smoothed version of well-log data. Well-log data refer to a suite of physical measurements acquired inside wells using various logging tools. These logs provide a detailed record of the geological formations the well has penetrated, offering crucial information such as the rock's seismic velocity, density, resistivity, and natural radiation, among others. Well-log data and seismic data differ significantly in scale and resolution: well-logs are high-frequency records with data points captured at intervals as small as a few centimeters, whereas seismic data have lower frequencies, typically ranging from 10 Hz to 100 Hz, corresponding to a resolution measured in meters.

\textbf{Multiresolution hash encoding:} 
Encoding input coordinates into a higher-dimensional space via pre-defined Fourier feature embeddings (or Gaussian-based alternatives) has been shown to significantly increase the ability of coordinate-based learning to capture high-frequency information in the sought-after solution. Results are reported across a variety of computer vision tasks \cite{tancik_fourier_2020, mildenhall_nerf_2020, zheng_rethinking_2021, sun_coil_2021}. Notably,~\cite{muller_instant_2022} introduced multiresolution hash encoding, a sparse grid-based parametric embedding that outperformed state-of-the-art non-parametric frequency-based encodings. This method utilizes a multi-resolution hash table with trainable feature vectors that are optimized through backpropagation, alongside the subsequent neural network parameters. This method leverages spatial hashing, where coordinates serve as the keys for a hash function, facilitating the retrieval of hash table indices. The hash function $h$ for a coordinate point $\mathbf{x}$ is defined as:
\begin{equation}
    h(\mathbf{x})=\left(\bigoplus_{i=1}^3 x_i \pi_i\right) \bmod T,
\end{equation}  
where $\bigoplus$ denotes the binary XOR operation, and $\pi_i$ represents a large prime number chosen by the user. This function computes the indices for a hash table of size $T$ that contains feature vectors of dimension $F$ at every entry. Given its multi-resolution nature, this operation is performed at multiple predefined resolutions. Each input coordinate is scaled to a specific resolution, and through floor and ceiling operations, the surrounding points' corners at a given resolution are ascertained—yielding four corner points for a 2D problem and eight for a 3D problem. These corners are fed into the hash function, with the resultant feature vectors being interpolated to obtain the corresponding feature vector at the input point.

\textbf{Neural Network Architecture:} The neural network component of IntraSeismic is a two-layer MLP network with ReLU activation function. Each layer contains 64 neurons. 

\textbf{Modeling Operator:} In the applications presented in subsequent sections, we utilize the post-stack seismic modeling operator, as defined in Equation~\ref{eqn:linear_poststack}.

\textbf{Optimizer:} We use the Adam optimizer \cite{kingma_adam_2017} with a learning rate of $10^{-3}$ to solve for the IntraSeismic parameters from the MLP network and hash encoding tables.

\textbf{Implementation details:}
The IntraSeismic framework is fully implemented in PyTorch 2.0.1, and all experiments are conducted on an AMD EPYC 7713 64-Core Processor equipped with a single NVIDIA TESLA A100. The modeling operator is implemented and wrapped into PyTorch using the PyLops library \cite{Ravasi2020}.

\section{Experiments and results}
    \label{experiments}
\subsection{Synthetic data}
\begin{figure*}
    \centering
    \includegraphics[width=\textwidth]{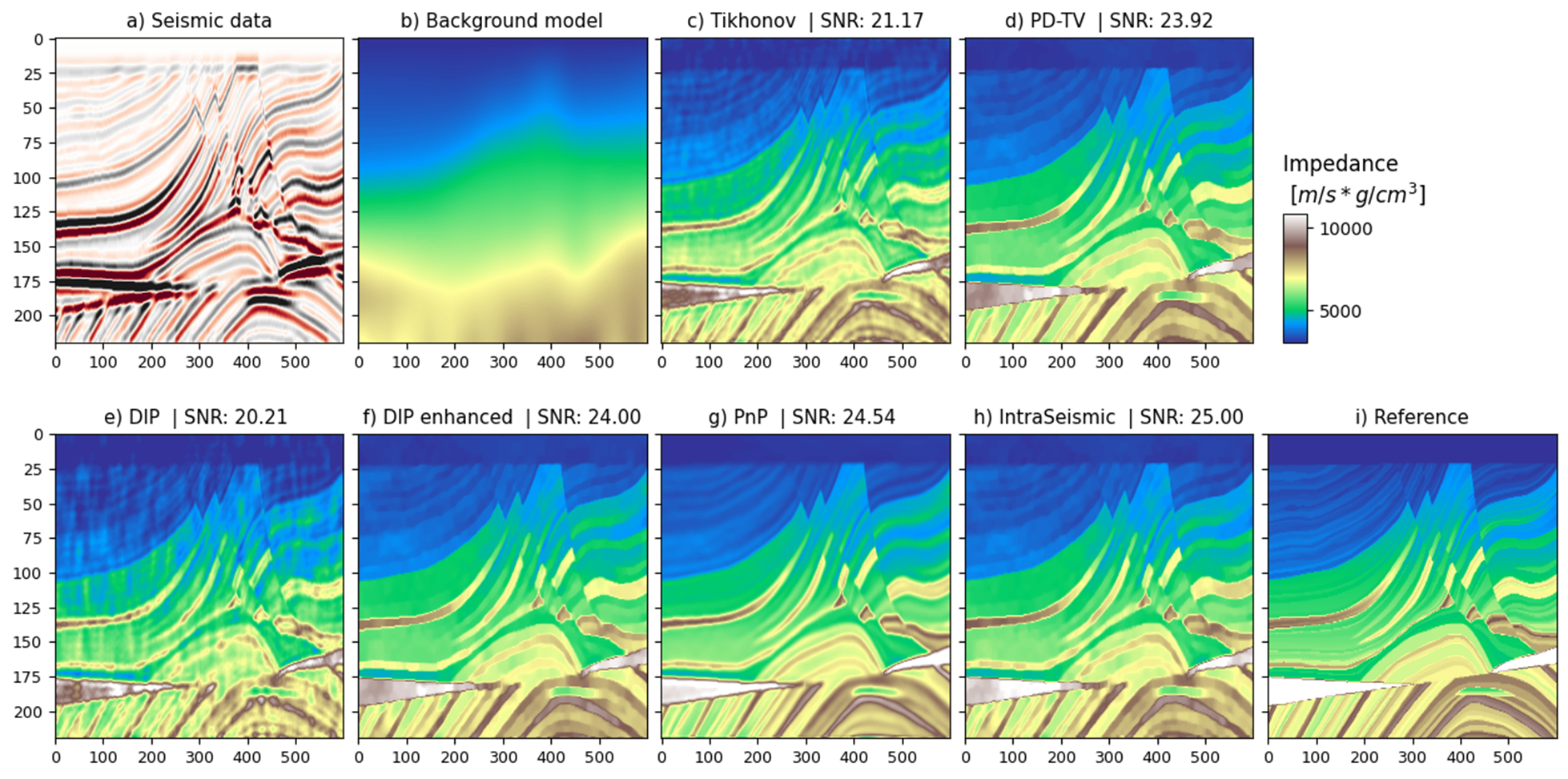}
    \caption{Inversion of the synthetic 2D Marmousi dataset with a noise level of $\sigma=0.1$: a) seismic data, b) background model, c) Tikhonov regularization, d) TV-regularization using the Primal-Dual solver, e) DIP, f) DIP with additional priors, g) PnP, h) IntraSeismic, and i) ground truth.}
    \label{fig:marm_results}
\end{figure*}

\begin{figure*}
    \centering
    \includegraphics[width=\textwidth]{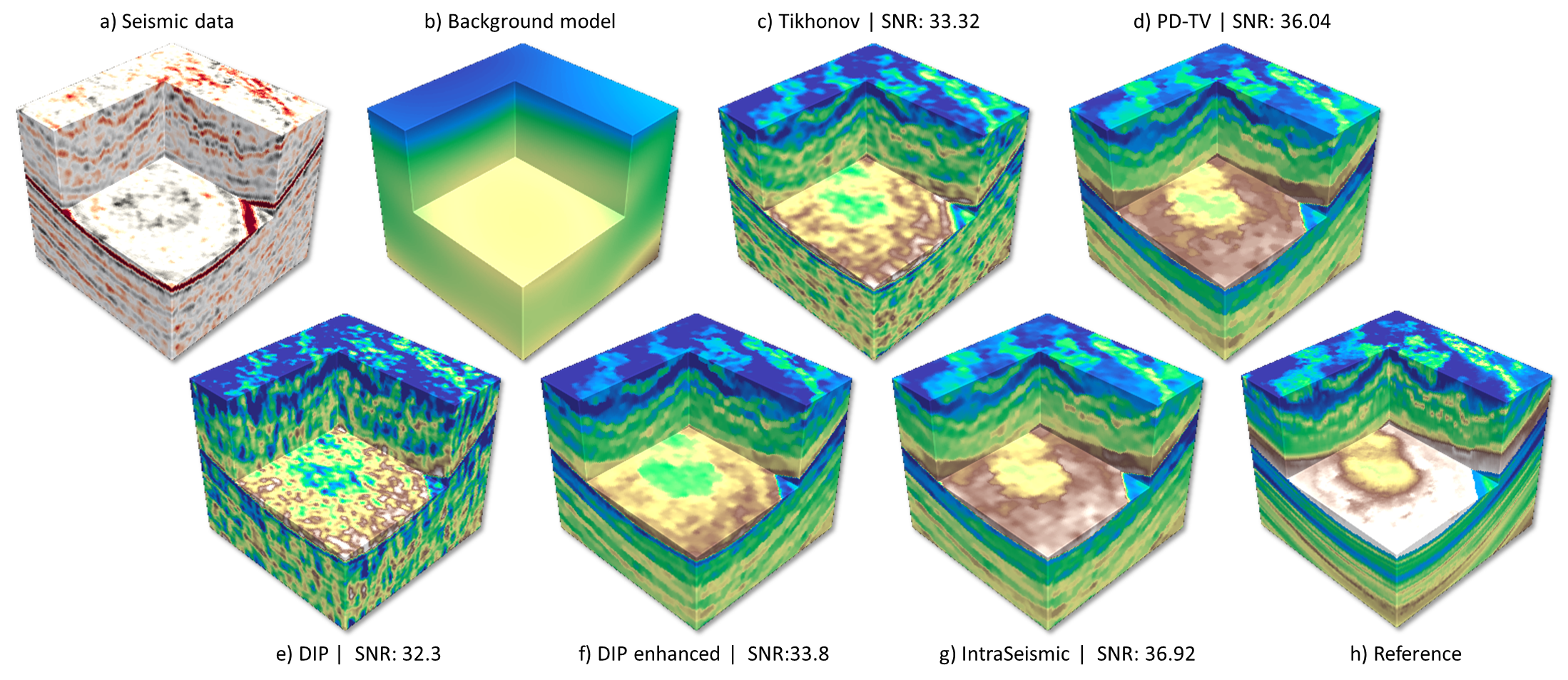}
    \caption{Inversion of the synthetic 3D SEAM dataset: a) seismic data, b) background model, c) Tikhonov regularization, d) TV-regularization using the Primal-Dual solver, e) DIP, f) DIP with additional priors, g) IntraSeismic and h) ground truth.}
    \label{fig:seam_results}
\end{figure*}

To assess the effectiveness and performance of IntraSeismic across different scenarios, we apply the proposed algorithm to both 2D and 3D synthetic post-stack seismic datasets, incorporating band-pass Gaussian noise at varying standard deviations: noise-free ($\sigma = 0$), $\sigma = 0.1$, and $\sigma = 0.2$. It is worth noting that the seismic dataset is scaled from $-1$ to $1$. In Table~\ref{tab:snr_comparison}, we report both the Signal-to-Noise Ratio (SNR) of the obtained solution and the number of iterations needed to achieve such a result. In the subsequent sections, we illustrate the results when using a moderate noise level ($\sigma = 0.1$), which is reflective of the usual noise levels encountered in field data. Figures~\ref{fig:marm_results}a and \ref{fig:seam_results}a display the corresponding data for the Marmousi and SEAM models. In both cases, the seismic data is modeled using a Ricker wavelet~\cite{ricker_further_1943} with a peak frequency of 15 Hz, and the background models are smoothed versions of the original acoustic impedance models (Figures~\ref{fig:marm_results}b, and \ref{fig:seam_results}b).

\begin{table*}[htbp]
  \centering 
  \caption{Inversion results for IntraSeismic and benchmark methods for the Marmousi and SEAM datasets and different noise levels. T SNR /\g{Iterations} are shown in each entry, and \textbf{bold} is used to indicate the best performing algorithm for any combination of model and noise.}.
  \label{tab:snr_comparison}
  \begin{footnotesize}
  \begin{tabular}{
    l 
    S[table-format=2.2] 
    S[table-format=2.2]
    S[table-format=2.2]
    S[table-format=2.2]
    S[table-format=2.2]
    S[table-format=2.2]}
    \toprule
    & \multicolumn{3}{c}{Marmousi} & \multicolumn{3}{c}{SEAM} \\
    \cmidrule(lr){2-4} \cmidrule(lr){5-7} 
    {Method} & {$\sigma=0$} & {$\sigma=0.1$} & {$\sigma=0.2$} & {$\sigma=0$} & {$\sigma=0.1$} & {$\sigma=0.2$} \\
    \midrule
    Tikhonov \cite{romero_plug_2022} & {21.50 \g{/71}} & {21.17 \g{/69}} & {20.35 \g{/69}} & {34.11 \g{/352}} & {33.32 \g{/400}} & {31.53 \g{/332}} \\
    PD-TV \cite{romero_plug_2022} & {24.74 \g{/4000}} & {23.92 \g{/38.70}} & {22.37 \g{/3580}} & {36.9 \g{/4000}} & {36.01 \g{/3000}} & {34.50 \g{/2000}} \\
    DIP \cite{ravasi_deep_2023} & {22.19 \g{/1475}} & {20.21 \g{/590}} & {17.61 \g{/306}} & {36.1 \g{/7000}} & {33.4 \g{/230}} & {28.00 \g{/7000}} \\
    DIP enhanced & {24.64 \g{/1238}} & {24.00 \g{/407}} & {23.71 \g{/349}} & {38.44 \g{/4100}} & {34.6 \g{/4600}} & {32.66 \g{/2600}} \\
    PnP \cite{romero_plug_2022} & {25.25 \g{/1280}} & {24.54 \g{/1660}} & {\textbf{23.37} \g{/1230}} & {N/A} & {N/A} & {N/A} \\
    IntraSeismic & {\textbf{27.34} \g{/1500}} & {\textbf{25.00} \g{/457}} & {22.93 \g{/437}} & {\textbf{40.84} \g{/393}} & {\textbf{36.92} \g{/217}} & {\textbf{34.5} \g{/307} }\\
    \bottomrule
  \end{tabular}
  \end{footnotesize}
\end{table*}

\indent The IntraSeismic result on synthetic data is benchmarked with two variational approaches, Tikhonov regularization (the standard method used in industry) and total-variation regularization (TV) using the Primal-Dual solver \cite{Ravasi2022}. Additionally, we compare IntraSeismic with the state-of-the-art hybrid methods proposed in the literature for post-stack seismic inversion, namely Deep Image Prior~\cite{lipari_post-stack_2021} and Plug-and-Play priors with deep denoisers \cite{romero_plug_2022}. 

\begin{figure}[t]
  \centering
   \includegraphics[width=0.5\linewidth]{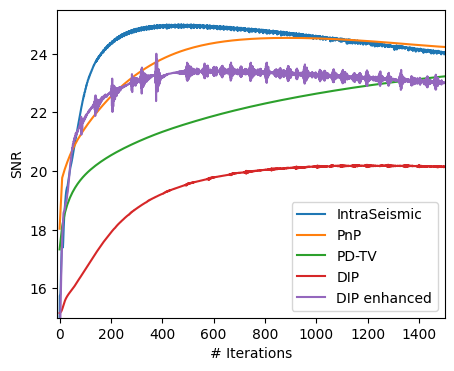}
   \caption{SNR evolution through iterations for IntraSeismic and benchmark methods in the inversion of the Marmousi data.}
   \vspace{-0.3cm}
   \label{fig:marm_snr}
\end{figure}

\indent For the PnP results, we utilized the published code \cite{romero_plug_2022}. This code employs the DRUNet denoiser \cite{zhang_plug-and-play_2022} within a Primal-Dual framework, functioning as a regularizer which achieved state-of-the-art outcomes in 2D post-stack seismic inversion. As of the publication date of our paper, we are not aware of any published implementation of PnP for 3D data. Consequently, we do not include a comparison with PnP in our 3D synthetic and field data experiments. In the case of DIP, \cite{ravasi_deep_2023} showed that the originally proposed method becomes sub-optimal in the presence of noise in post-stack seismic inversion. This limitation may stem from the absence of prior information that guides the inversion toward geologically consistent solutions. In our study, we further modify DIP by incorporating the same regularization terms as those used in IntraSeismic (see equation~\ref{eqn:intraseismic}), and we refer to it as DIP enhanced. We believe that this facilitates a more equitable comparison between the DIP and IntraSeismic methods, where their main difference lies in the parameterization of the acoustic impedance model (i.e., a CNN that warps a noise vector into the sought-after solution in DIP, and an MLP that transforms coordinates into values of the solution in IntraSeismic). Finally, we decided not to provide any comparison with state-of-the-art supervised learning approaches as that would introduce additional factors (e.g., size of the training dataset) that greatly affect the quality of the final solution. \\
\newline
\textbf{Marmousi model} \\
The Marmousi model \cite{brougois_marmousi_1990} is a synthetically created, realistic 2D subsurface velocity model that emulates complex geological structures with sharp velocity contrasts (see Figure \ref{fig:marm_results}i). The Marmousi model is the most widely used model in geophysics, thereby serving as a benchmark for various seismic imaging algorithms. Figures~\ref{fig:marm_results}(c to f) illustrate the inversion results for the Marmousi dataset at a noise level of $\sigma=0.1$, comparing standard variational methods, hybrid algorithms, and IntraSeismic. IntraSeismic retrieves a high-resolution acoustic impedance image of the Marmousi model, where discontinuities and layer contrasts are clearly defined.
As indicated in Table \ref{tab:snr_comparison}, IntraSeismic achieves the highest SNR in the Marmousi model for both noise-free conditions and noisy data with $\sigma=0.1$, and it is slightly surpassed by PnP for the highest noise level ($\sigma=0.2$). Moreover, Table \ref{tab:snr_comparison} and Figure~\ref{fig:marm_snr} highlight the fast convergence of IntraSeismic, reaching the highest SNR in less than $500$ iterations for noisy data.\\

\begin{figure}[!]
  \centering
   \includegraphics[width=0.5\linewidth]{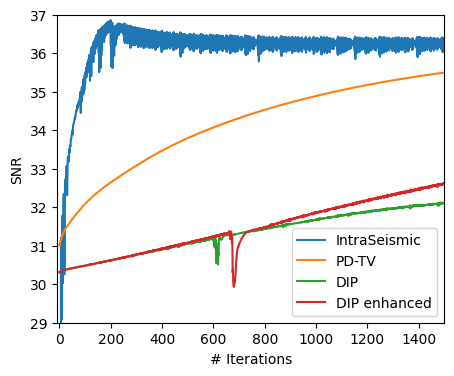}
   \caption{SNR evolution through iterations for IntraSeismic and benchmark methods in the inversion of the SEAM data.}
   \vspace{-0.2cm}
   \label{fig:seam_snr}
\end{figure}

\noindent{\textbf{SEAM model}}\\
The model employed in this study is a volume of size $180 \times 180 \times 180$, extracted from the SEAM Phase I velocity model created by the SEG Advanced Modeling Consortium \cite{Fehler2011}. The SEAM model showcases realistic geological heterogeneities and discontinuities across all three spatial dimensions and exhibits high contrasts between some of its layers (Figure~\ref{fig:seam_results}g). Evidently, the IntraSeismic method (Figure~\ref{fig:seam_results}f) produces the solution with the highest SNR, leading to a more precise 3D reconstruction of the SEAM model. As demonstrated in Table~\ref{tab:snr_comparison}, IntraSeismic achieves the highest SNR across all noise levels for the SEAM dataset. Moreover, it converges more rapidly than existing benchmark methods, requiring fewer than 400 iterations in every evaluated noise scenario. This is illustrated in Figure~\ref{fig:seam_snr}, where the SNR evolution over iterations is shown for some of the benchmark methods and IntraSeismic for the case of noisy data with $\sigma=0.1$.

\subsection{Field data: Sleipner dataset}

\begin{figure*}
    \centering
    \includegraphics[width=\textwidth]{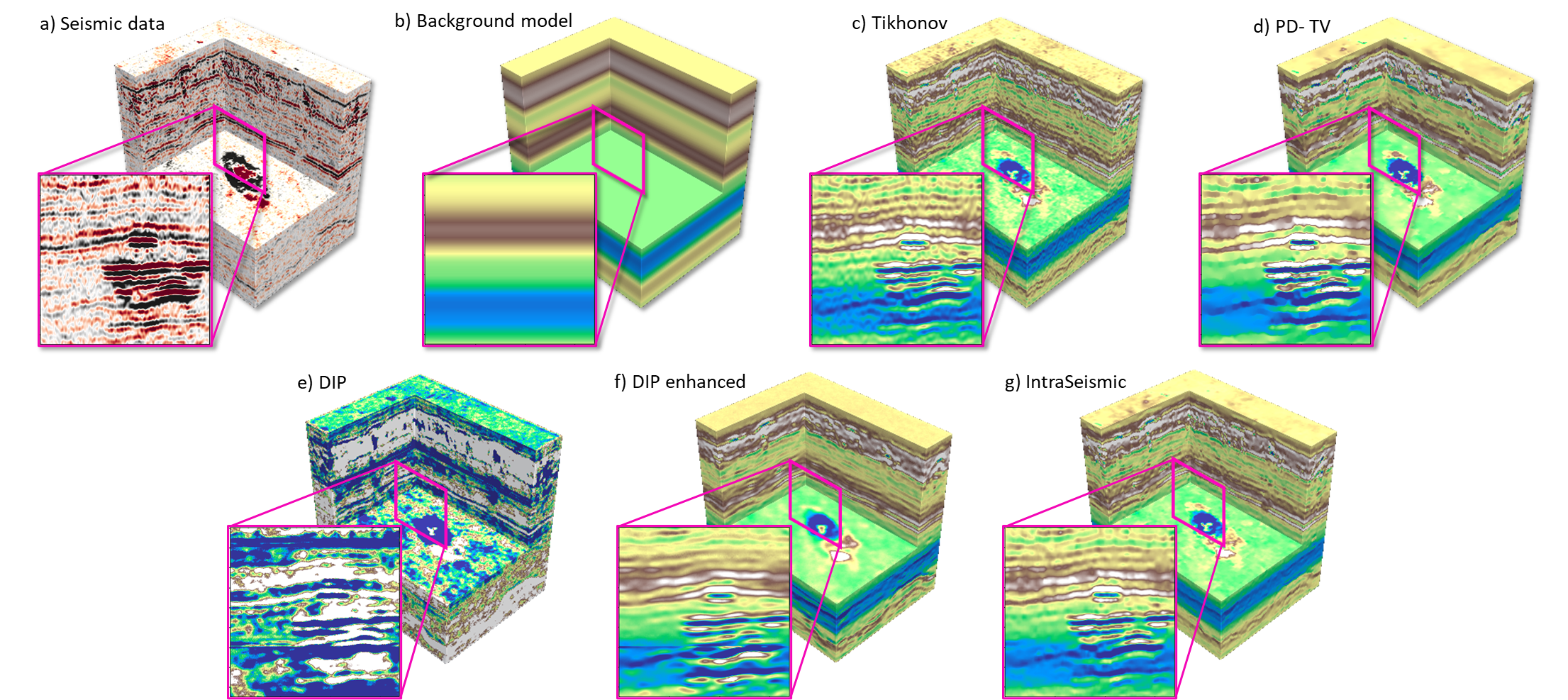}
    \caption{Inversion of Sleipner dataset using IntraSeismic and benchmark methods: a) seismic data, b) background model, c) Tikhonov regularization, d) TV-regularization using the Primal-Dual solver, e) DIP, f) DIP with additional priors, and g) IntraSeismic.}
    \label{fig:sleip_results}
\end{figure*}

 The Sleipner seismic dataset was acquired in 2001 in the North Sea as part of a seismic monitoring program of an ongoing carbon capture and storage project in the Sleipner field \cite{furre2017}. We use a subvolume of size $300 \times 300 \times 250$  (Figure~\ref{fig:sleip_results}a), which exhibits a predominantly horizontal geology disrupted by strong contrast values in the CO$_2$ reservoir. The statistical wavelet and the background model (Figure~\ref{fig:sleip_results}b) are taken from \cite{Romero2023}. Figure~\ref{fig:sleip_results} shows the inversion results for the benchmark methods and IntraSeismic. In this case, we do not observe any major scale difference between the IntraSeismic and PD-TV results; however, when looking at the inversion results in detail, IntraSeismic shows smoother transitions in the horizontal direction (a sought-after characteristic in geology) and heightened resolution in the vertical direction that accentuates the layer contrasts. After 200 iterations (spanning 40 seconds), we notice minimal changes in the model update, which we interpret as the convergence point.

\subsection{Uncertainty quantification}
    \label{section_uq}
\begin{figure}[h]
  \centering
   \includegraphics[width=0.5\linewidth]{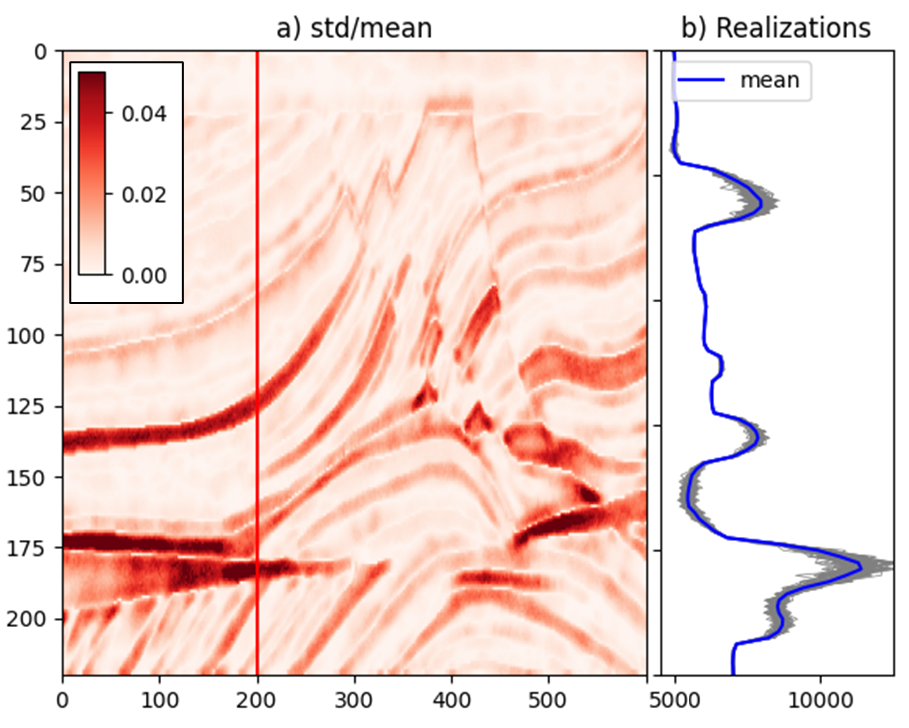}
   \caption{Monte Carlo Dropout uncertainty quantification result for Marmousi model: a) std/mean, the red line indicates the extracted column shown in b).}
   \vspace{-0.3cm}
   \label{fig:marm_uq}
\end{figure}

Uncertainty quantification (UQ) is pivotal in any seismic inversion process given the non-uniqueness of the seismic inverse problem, the inherent complexity and variability of the Earth's subsurface, and the previously discussed limitations of the observed data. As such, UQ offers insight into the range of possible subsurface models that can equally fit the observed data. This knowledge is instrumental for informed decision-making in scenarios with significant financial and environmental implications \cite{smith2013uq}. Furthermore, understanding the uncertainties associated with a seismic inversion product can guide subsequent data acquisition campaigns to refine results and provide a clearer picture of result reliability and associated risks. When a neural network is involved directly (via supervised learning) or indirectly (as a regularizer or preconditioner) in the solution of the seismic inverse problem, additional uncertainty is introduced in the process. Among other techniques to quantify such an uncertainty component, Monte Carlo Dropout (MCD) has been shown to be an effective and easy-to-implement method to quantify the so-called epistemic uncertainty \cite{gal_dropout_2016}. By retaining the dropout layers during the inference phase, rather than just during training, and performing multiple evaluations of the network with different neurons "dropped out" each time, one can obtain samples of an approximate posterior distribution \cite{kendall_what_2017}. We propose to add dropout layers to IntraSeismic's MLP network and to perform MCD sampling once the optimization process is completed. The proposed technique is applied to the Marmousi model. More specifically, in this case, we apply dropout to both of the layers of the MLP network (Figure \ref{fig:arch}) with a dropout probability of $0.2$, and we compute a total of $100$ realizations, which are used to evaluate sample statistics (e.g., sample mean and standard deviation). Figure~\ref{fig:marm_uq} corresponds to the obtained standard deviation divided by the mean of the Marmousi model parameters that shows the IntraSeismic's confidence levels and highlights areas in the model with potential ambiguity in the solution. The layers with higher impedance values exhibit the highest uncertainty, possibly due to TV regularization, which might lead to a diminished contrast. 

\subsection{Adding hard-constraints}
\begin{figure}[h]
  \centering
   \includegraphics[width=0.7\linewidth]{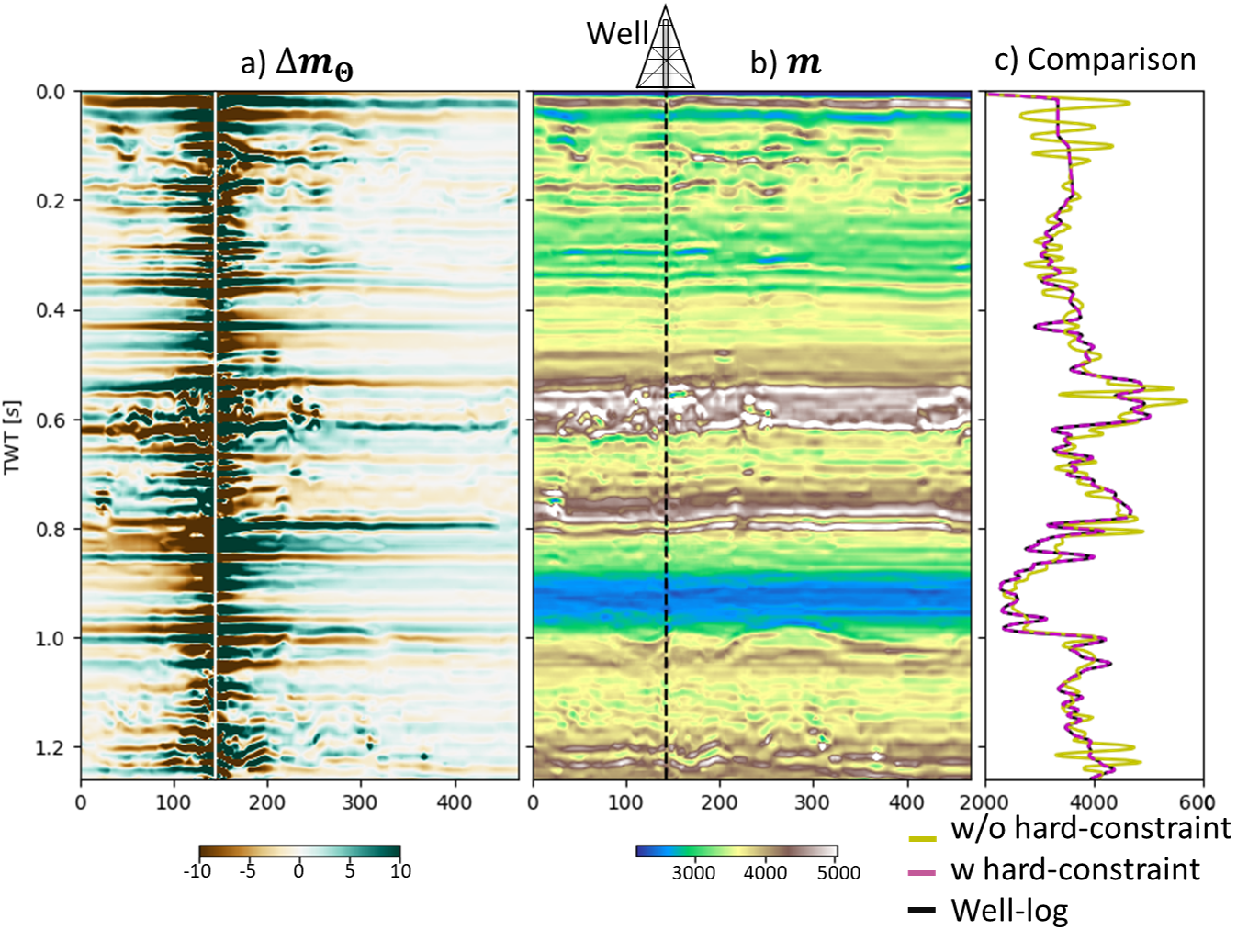}
   \vspace{-0.5cm}
   \caption{Hard-constrained inversion of Sleipner dataset: a) Output of the network $\Delta \boldsymbol{m}_{\Theta}$, b) final estimated acoustic impedance model and c) comparison at the well location of the inversion with and without hard-contraint.}
   \label{fig:sleip_hc}
   \vspace{-0.4cm}
\end{figure}

Well-log data provide the closest approximation to the ground truth in seismic inversion, as they are actual recordings of subsurface physical properties. As such, they can serve as critical anchor points, ensuring that the seismic inversion results are consistent with real-world measurements at specific locations. One way to incorporate well-log data in the seismic inversion process is by integrating it directly into the original optimization framework. This can be achieved by introducing a soft constraint to the optimization problem as formulated in equation \ref{eqn:poststack_inv}. Specifically, the term $ \gamma\left\|\boldsymbol{m}_i-\boldsymbol{m}_{\text{well}}\right\|_2^2$ can be added, where $\gamma$ is a weighting factor that balances the influence of the well-log data, $\mathbf{m_i}$ represents the model parameters at the well position, and $\boldsymbol{m}_{\text{well}}$ denotes the measurements from the well-log. This addition ensures that the inversion solution aligns more closely with the well-log observations.
An alternative for incorporating well-log data into the seismic inversion process involves the application of a hard constraint, which enforces an exact match between the impedance model and the well-log data ($\boldsymbol{m}_i=\boldsymbol{m}_{\text{well}}$). This approach is particularly advantageous in instances of highly noisy data where the integrity of well data is prioritized or in scenarios with numerous wells within a development field, where the objective is to construct a robust model that accurately reflects all available well data. In IntraSeismic, the integration of well data can be naturally achieved by redefining the network's output as follows \cite{schiassi_extreme_2020}:
\begin{equation}
    \boldsymbol{m} = (\boldsymbol{x} - \boldsymbol{x}_{\text{well}}) \Delta \boldsymbol{m}_{\Theta} + \boldsymbol{m}_{\text{well}}
\end{equation}
where $\boldsymbol{x}$ represents the coordinate vector, while $\boldsymbol{x}_{\text{well}}$ and $\boldsymbol{m}_{\text{well}}$ denote the coordinates and values of the well-log data, respectively. The term $\Delta \boldsymbol{m}_{\Theta}$ corresponds to the output of the neural network. Within this framework, the neural network's output ensures compliance with the hard constraint and acts as an update to the well data away from its position. This approach is implemented for the Sleipner dataset, forcing the solution to match the available acoustic impedance well-log in the middle of the model (Figure \ref{fig:sleip_hc}).

\subsection{Irregular data access and data compression}
Geoscientists are constantly tasked to analyze vast amounts of seismic data, a process usually referred to as seismic interpretation \cite{Herron}. As such, the property models produced by different seismic inversion algorithms are commonly visualized along different axes (i.e., inline, crossline, time/depth) or along irregular surfaces corresponding to key geological horizons. Conventional seismic interpretation software packages perform such extraction procedures operating directly with data stored on fast-access file storage systems or on local memory when the data size allows it. \\
\indent However, a user must select upfront the layout of the data, placing the most commonly accessed direction over the fast axis and the least commonly accessed direction over the slow one or replicating the same information over multiple files with different layouts. In order to reduce the amount of I/O, the seismic industry has recently placed tremendous focus on compression algorithms, with the popular ZFP data format \cite{lindstrom2014} being widely adopted in both proprietary and open-source data formats (e.g., MDIO \cite{sansal_mdio_2023}). A second by-product of the IntraSeismic framework is represented by the coordinate-based parameterization of the sought-after acoustic impedance model. As such, the cost of accessing the model values at $N$ (possibly random) coordinates is independent of the location and order of such points. We believe that this unique feature of IntraSeismic could revolutionize how seismic interpretation software packages store and access seismic volumes.\\
\indent Moreover, given that the complete knowledge of the acoustic impedance model is encapsulated within the learned hash tables and the parameters of the MLP, the IntraSeismic framework can be considered an effective data compression tool. The SEAM model, illustrated in Figure~\ref{fig:seam_results}, comprises of $5.83$ million data points, yet the IntraSeismic framework can represent it with only $1.16$ million parameters. This yields a compression ratio of approximately $5$. In contrast, the ZFP compression algorithm achieves a compression ratio of approximately $15$ for the SEAM model, with an error margin of $0.1\%$. A significant factor in the high compression efficiency of ZFP can be attributed to its quantization step performed prior to storing the compressed data. With quantization emerging as a focal point of interest within the field of Deep Learning —for its utility in representing neural network weights in a reduced and efficient format— it is plausible to anticipate an improvement in IntraSeismic’s compression ratio with the application of quantization to the hash table and MLP parameters.

\section{Conclusions and future directions}
In this study, we have introduced the IntraSeimic framework, a hybrid data-driven and model-driven strategy to solve seismic inverse problems. We presented and critically evaluated the IntraSeismic framework by applying it to both realistic synthetic and field post-stack seismic datasets, and benchmarked it against prevalent variational approaches and deep-learning-based methods in the seismic inversion domain. Our findings underscore IntraSeismic's superior capability in producing high-resolution impedance models and faster convergence. Additionally, the IntraSeismic framework's inherent coordinate-based parametrization offers revolutionary prospects for the seismic interpretation domain, especially in terms of data access and compression. The introduction of Monte Carlo Dropout to IntraSeismic further empowers it to quantify uncertainties, a critical aspect in seismic inversion. 

The visual similarity of the IntraSeismic results to those obtained using the PD-TV method stems mainly from the shared use of the TV regularization strategy. Both IntraSeismic and PD-TV exhibit challenges in recovering high impedance layers—a feat where the PnP method showed superiority in the Marmousi example. Such limitations might arise from selecting the regularization term itself, which can inadvertently constrain the framework's modeling capacity, leading to inherent trade-offs like the loss of contrast commonly associated with TV regularization. Therefore, future research should aim to explore learned regularization strategies (like PnP) within IntraSeismic-type frameworks. Finally, while this study has focused on the application of IntraSeismic to post-stack seismic inversion, the framework's design allows for straightforward adaptation to other inversion scenarios, such as pre-stack seismic inversion and image domain least-squares migration (additional details provided in the supplementary material).

\section*{Acknowledgments}
The authors thank King Abdullah University of Science \& Technology (KAUST) for supporting this research as well as Equinor and partners for releasing the 4D Sleipner dataset (available at \url{https://co2datashare.org/dataset/sleipner-2019-benchmark-model}).  For computer time, this research used the resources of the Supercomputing Laboratory at KAUST in Thuwal, Saudi Arabia.  

\bibliographystyle{unsrt}  
\bibliography{references}

\begin{thebibliography}{10}

\bibitem{kearey_2002}
Philip Kearey, Michael Brooks, and Ian Hill.
\newblock An {Introduction} to {Geophysical} {Exploration}, 3rd {Edition} {\textbar} {Wiley}, 2002.

\bibitem{faccio2020}
Daniele Faccio, Andreas Velten, and Gordon Wetzstein.
\newblock Non-line-of-sight imaging.
\newblock {\em Nature Reviews Physics}, 2:318–--327, 2020.

\bibitem{telford_applied_1990}
William~Murray Telford, Lloyd~P. Geldart, and Robert~E. Sheriff.
\newblock Applied {Geophysics}, October 1990.
\newblock ISBN: 9781139167932 Publisher: Cambridge University Press.

\bibitem{sheriff_exploration_1995}
Robert~E. Sheriff and Lloyd~P. Geldart.
\newblock {\em Exploration {Seismology}}.
\newblock Cambridge University Press, Cambridge, 2 edition, 1995.

\bibitem{Yilmaz}
Öz Yilmaz.
\newblock {\em Seismic data analysis}.
\newblock Society of Exploration Geophysicists, 2001.

\bibitem{oz_engineering_2015}
Öz Yilmaz and Richard~D. Miller.
\newblock {\em Engineering {Seismology}: {With} {Applications} to {Geotechnical} {Engineering}}.
\newblock Society of Exploration Geophysicists, January 2015.

\bibitem{steeples_seismic_1988}
Don~W. Steeples and Richard~D. Miller.
\newblock Seismic {Reflection} {Methods} {Applied} {To} {Engineering}, {Environmental}, {And} {Ground}-{Water} {Problems}.
\newblock page~cp. European Association of Geoscientists \& Engineers, March 1988.
\newblock ISSN: 2214-4609.

\bibitem{lumley_4d_2010}
David Lumley.
\newblock {4D} seismic monitoring of {CO2} sequestration.
\newblock {\em The Leading Edge}, 29(2):150--155, February 2010.
\newblock Publisher: Society of Exploration Geophysicists.

\bibitem{tarantola_inversion_1984}
Albert Tarantola.
\newblock Inversion of seismic reflection data in the acoustic approximation.
\newblock {\em GEOPHYSICS}, 49(8):1259--1266, August 1984.
\newblock Publisher: Society of Exploration Geophysicists.

\bibitem{Tikhonov1977}
Andrey~N. Tikhonov and Vasiliy~.Y. Arsenin.
\newblock {\em Solution of Ill-posed Problems}.
\newblock Washington: Winston \& Sons, 1977.

\bibitem{rudin_nonlinear_1992}
Leonid~I. Rudin, Stanley Osher, and Emad Fatemi.
\newblock Nonlinear total variation based noise removal algorithms.
\newblock {\em Physica D: Nonlinear Phenomena}, 60(1):259--268, November 1992.

\bibitem{Gholami}
Ali Gholami.
\newblock Nonlinear multichannel impedance inversion by total-variation regularization.
\newblock {\em Geophysics}, 80:R217--R224, 2015.

\bibitem{Wang}
Dehua Wang, Jinghuai Gao, and Hongan Zhou.
\newblock Data-driven multichannel seismic impedance inversion with anisotropic total variation regularization.
\newblock {\em Journal of Inverse and Ill-Posed Problems}, 2018(2):229–241, 2019.

\bibitem{biswas_prestack_2019}
Reetam Biswas, Mrinal~K. Sen, Vishal Das, and Tapan Mukerji.
\newblock Prestack and poststack inversion using a physics-guided convolutional neural network.
\newblock {\em Interpretation}, 7(3), August 2019.
\newblock Publisher: Society of Exploration Geophysicists.

\bibitem{das_convolutional_2019}
Vishal Das, Ahinoam Pollack, Uri Wollner, and Tapan Mukerji.
\newblock Convolutional neural network for seismic impedance inversion.
\newblock {\em GEOPHYSICS}, 84(6):R869--R880, November 2019.
\newblock Publisher: Society of Exploration Geophysicists.

\bibitem{smith_robust_2022}
Robert Smith, Philippe Nivlet, Hussain Alfayez, and Nasher AlBinHassan.
\newblock Robust deep learning-based seismic inversion workflow using temporal convolutional networks.
\newblock {\em Interpretation}, 10(2):SC41--SC55, May 2022.
\newblock Publisher: Society of Exploration Geophysicists.

\bibitem{wu_seismic_2022}
Bangyu Wu, Qiao Xie, and Baohai Wu.
\newblock Seismic {Impedance} {Inversion} {Based} on {Residual} {Attention} {Network}.
\newblock {\em IEEE Transactions on Geoscience and Remote Sensing}, 60:1--17, 2022.
\newblock Conference Name: IEEE Transactions on Geoscience and Remote Sensing.

\bibitem{ravasi_deep_2023}
Matteo Ravasi, Nick Luiken, Juan Romero, and Miguel Corrales.
\newblock Deep {Learning} to replace or augment model-based seismic inversion?
\newblock volume 2023, pages 1--5. European Association of Geoscientists \& Engineers, June 2023.
\newblock ISSN: 2214-4609 Issue: 1.

\bibitem{ulyanov_deep_2020}
Dmitry Ulyanov, Andrea Vedaldi, and Victor Lempitsky.
\newblock Deep {Image} {Prior}.
\newblock {\em International Journal of Computer Vision}, 128(7):1867--1888, July 2020.
\newblock arXiv:1711.10925 [cs, stat].

\bibitem{lipari_post-stack_2021}
Vicenzo Lipari, Francesco Picetti, P.~Bestagini, and Stefano Tubaro.
\newblock Post-{Stack} {Inversion} with {Uncertainty} {Estimation} through {Bayesian} {Deep} {Image} {Prior}.
\newblock volume 2021, pages 1--5. European Association of Geoscientists \& Engineers, October 2021.
\newblock ISSN: 2214-4609 Issue: 1.

\bibitem{venkatakrishnan_plug-and-play_2013}
Singanallur~V. Venkatakrishnan, Charles~A. Bouman, and Brendt Wohlberg.
\newblock Plug-and-{Play} priors for model based reconstruction.
\newblock In {\em 2013 {IEEE} {Global} {Conference} on {Signal} and {Information} {Processing}}, pages 945--948, December 2013.

\bibitem{zhang_plug-and-play_2022}
Kai Zhang, Yawei Li, Wangmeng Zuo, Lei Zhang, Luc Van~Gool, and Radu Timofte.
\newblock Plug-and-{Play} {Image} {Restoration} {With} {Deep} {Denoiser} {Prior}.
\newblock {\em IEEE Transactions on Pattern Analysis and Machine Intelligence}, 44(10):6360--6376, October 2022.
\newblock Conference Name: IEEE Transactions on Pattern Analysis and Machine Intelligence.

\bibitem{romero_plug_2022}
Juan Romero, Miguel Corrales, Nick Luiken, and Matteo Ravasi.
\newblock Plug and {Play} {Post}-{Stack} {Seismic} {Inversion} with {CNN}-{Based} {Denoisers}.
\newblock volume 2022, pages 1--5. European Association of Geoscientists \& Engineers, October 2022.
\newblock ISSN: 2214-4609 Issue: 1.

\bibitem{ravasi_chapter_2023}
Matteo Ravasi, Juan Romero, Miguel Corrales, Nick Luiken, and Claire Birnie.
\newblock Chapter {Six} - {Striking} a balance: {Seismic} inversion with model- and data-driven priors.
\newblock In Shib~Sankar Ganguli and Vijay~Prasad Dimri, editors, {\em Developments in {Structural} {Geology} and {Tectonics}}, volume~6 of {\em Reservoir {Characterization}, {Modeling}, and {Quantitative} {Interpretation}}, pages 153--200. Elsevier, January 2023.

\bibitem{brougois_marmousi_1990}
Aline Brougois, Marielle Bourget, Patrick Lailly, Michel Poulet, Patrice Ricarte, and Roelof Versteeg.
\newblock Marmousi, model and data.
\newblock page~cp. European Association of Geoscientists \& Engineers, May 1990.
\newblock ISSN: 2214-4609.

\bibitem{Fehler2011}
Michael Fehler and P.~Joseph Keliher.
\newblock Seam phase 1: Challenges of subsalt imaging in tertiary basins, with emphasis on deepwater gulf of mexico.
\newblock Society of Exploration Geophysicists, 2011.

\bibitem{furre2017}
Anne-Kari Furre, Ola Eiken, Håvard Alnes, Jonas~Nesland Vevatne, and Anders~Fredrik Kiær.
\newblock 20 years of monitoring co2-injection at sleipner.
\newblock {\em Energy Procedia}, 114:3916--3926, 2017.
\newblock 13th International Conference on Greenhouse Gas Control Technologies, GHGT-13, 14-18 November 2016, Lausanne, Switzerland.

\bibitem{goupillaud_approach_1961}
Pierre~L. Goupillaud.
\newblock An approach to inverse filtering of near‐surface layer effects from seismic records.
\newblock {\em GEOPHYSICS}, 26(6):754--760, December 1961.
\newblock Publisher: Society of Exploration Geophysicists.

\bibitem{Stolt1985}
R.H. Stolt and A.B. Weglein.
\newblock Migration and inversion of seismic data.
\newblock {\em Geophysics}, 50:2458--2472, 1985.

\bibitem{Oldenburg1983}
D.W. Oldebnbur, T.~Scheuer, and S.~Levy.
\newblock Recovery of the acoustic impedance from reflection seismograms.
\newblock {\em Geophysics}, 48:1318--1337, 1983.

\bibitem{Russell1991}
Brian Russell and Dan Hampson.
\newblock Comparison of poststack seismic inversion methods.
\newblock {\em 61st Annual International Meeting, SEG}, pages 876--878, 1991.

\bibitem{tancik_fourier_2020}
Matthew Tancik, Pratul~P. Srinivasan, Ben Mildenhall, Sara Fridovich-Keil, Nithin Raghavan, Utkarsh Singhal, Ravi Ramamoorthi, Jonathan~T. Barron, and Ren Ng.
\newblock Fourier {Features} {Let} {Networks} {Learn} {High} {Frequency} {Functions} in {Low} {Dimensional} {Domains}, June 2020.
\newblock arXiv:2006.10739 [cs].

\bibitem{stanley_compositional_2007}
Kenneth~O. Stanley.
\newblock Compositional pattern producing networks: {A} novel abstraction of development.
\newblock {\em Genetic Programming and Evolvable Machines}, 8(2):131--162, June 2007.

\bibitem{chen_learning_2019}
Zhiqin Chen and Hao Zhang.
\newblock Learning {Implicit} {Fields} for {Generative} {Shape} {Modeling}, September 2019.
\newblock arXiv:1812.02822 [cs].

\bibitem{mildenhall_nerf_2020}
Ben Mildenhall, Pratul~P. Srinivasan, Matthew Tancik, Jonathan~T. Barron, Ravi Ramamoorthi, and Ren Ng.
\newblock {NeRF}: {Representing} {Scenes} as {Neural} {Radiance} {Fields} for {View} {Synthesis}, August 2020.
\newblock arXiv:2003.08934 [cs].

\bibitem{zang_intratomo_2021}
Guangming Zang, Ramzi Idoughi, Rui Li, Peter Wonka, and Wolfgang Heidrich.
\newblock {IntraTomo}: {Self}-supervised {Learning}-based {Tomography} via {Sinogram} {Synthesis} and {Prediction}.
\newblock In {\em 2021 {IEEE}/{CVF} {International} {Conference} on {Computer} {Vision} ({ICCV})}, pages 1940--1950, October 2021.
\newblock ISSN: 2380-7504.

\bibitem{sun2023}
Jian Sun, Kristopher Innanen, Tianze Zhang, and Daniel Trad.
\newblock Implicit seismic full waveform inversion with deep neural representation.
\newblock {\em Journal of Geophysical Research: Solid Earth}, 128(3):e2022JB025964, 2023.

\bibitem{muller_instant_2022}
Thomas Müller, Alex Evans, Christoph Schied, and Alexander Keller.
\newblock Instant {Neural} {Graphics} {Primitives} with a {Multiresolution} {Hash} {Encoding}.
\newblock {\em ACM Transactions on Graphics}, 41(4):1--15, July 2022.
\newblock arXiv:2201.05989 [cs].

\bibitem{zheng_rethinking_2021}
Jianqiao Zheng, Sameera Ramasinghe, and Simon Lucey.
\newblock Rethinking {Positional} {Encoding}, October 2021.
\newblock arXiv:2107.02561 [cs].

\bibitem{sun_coil_2021}
Yu~Sun, Jiaming Liu, Mingyang Xie, Brendt Wohlberg, and Ulugbek~S. Kamilov.
\newblock {CoIL}: {Coordinate}-based {Internal} {Learning} for {Imaging} {Inverse} {Problems}, February 2021.
\newblock arXiv:2102.05181 [eess].

\bibitem{kingma_adam_2017}
Diederik~P. Kingma and Jimmy Ba.
\newblock Adam: {A} {Method} for {Stochastic} {Optimization}, January 2014.
\newblock arXiv:1412.6980 [cs].

\bibitem{Ravasi2020}
Matteo Ravasi and Ivan Vasconcelos.
\newblock Pylops—a linear-operator python library for scalable algebra and optimization.
\newblock {\em SoftwareX}, 11:100361, 2020.

\bibitem{ricker_further_1943}
Norman Ricker.
\newblock Further developments in the wavelet theory of seismogram structure*.
\newblock {\em Bulletin of the Seismological Society of America}, 33(3):197--228, July 1943.

\bibitem{Ravasi2022}
Matteo Ravasi and Claire Birnie.
\newblock A joint inversion-segmentation approach to assisted seismic interpretation.
\newblock {\em Geophysical Journal International}, 28:893--912, 2022.

\bibitem{Romero2023}
Juan Romero, Nick Luiken, and Matteo Ravasi.
\newblock Seeing through the co2 plume: Joint inversion-segmentation of the sleipner 4d seismic data set.
\newblock {\em The Leading Edge}, 42:446–516, 2023.

\bibitem{smith2013uq}
Ralph~C. Smith.
\newblock {\em Uncertainty Quantification: Theory, Implementation, and Applications}.
\newblock Society for Industrial and Applied Mathematics, Philadelphia, PA, 2013.

\bibitem{gal_dropout_2016}
Yarin Gal and Zoubin Ghahramani.
\newblock Dropout as a {Bayesian} {Approximation}: {Representing} {Model} {Uncertainty} in {Deep} {Learning}.
\newblock In {\em Proceedings of {The} 33rd {International} {Conference} on {Machine} {Learning}}, pages 1050--1059. PMLR, June 2016.
\newblock ISSN: 1938-7228.

\bibitem{kendall_what_2017}
Alex Kendall and Yarin Gal.
\newblock What {Uncertainties} {Do} {We} {Need} in {Bayesian} {Deep} {Learning} for {Computer} {Vision}?, October 2017.
\newblock arXiv:1703.04977 [cs].

\bibitem{schiassi_extreme_2020}
Enrico Schiassi, Carl Leake, Mario De~Florio, Hunter Johnston, Roberto Furfaro, and Daniele Mortari.
\newblock Extreme {Theory} of {Functional} {Connections}: {A} {Physics}-{Informed} {Neural} {Network} {Method} for {Solving} {Parametric} {Differential} {Equations}, May 2020.
\newblock arXiv:2005.10632 [physics, stat].

\bibitem{Herron}
Donald~A. Herron.
\newblock {\em Seismic Intepretation}.
\newblock Society of Exploration Geophysicists, 2011.

\bibitem{lindstrom2014}
Peter Lindstrom.
\newblock Fixed-rate compressed floating-point arrays.
\newblock {\em IEEE Transactions on Visualization and Computer Graphics}, 20(12):2674--2683, 2014.

\bibitem{sansal_mdio_2023}
Altay Sansal, Sribharath Kainkaryam, Ben Lasscock, and Alejandro Valenciano.
\newblock {MDIO}: {Open}-source format for multidimensional energy data.
\newblock {\em The Leading Edge}, 42(7):465--473, July 2023.
\newblock Publisher: Society of Exploration Geophysicists.

\bibitem{aki2002}
Keiti Aki and Paul~G. Richards.
\newblock {\em Quantitative Seismology}.
\newblock W. H. Freeman and Company., 2002.

\bibitem{Zoeppritz1919}
K.~Zoeppritz.
\newblock Uber reflexion und durchgang seismischer wellen durch unstetigkeitsflächen.
\newblock {\em Nachrichten von der Königlichen Gesellschaft der Wissenschaften zu Göttingen, Mathematisch-physikalische Klasse}, page 66–84, 1919.

\bibitem{gary2006}
Gary~S. Martin, Robert Wiley, and Kurt~J. Marfurt.
\newblock {Marmousi2 : An elastic upgrade for Marmousi}.
\newblock {\em The Leading Edge}, 25(2):156--166, 02 2006.

\bibitem{Downton}
J~E Downton.
\newblock {\em Seismic parameter estimation from AVO inversion}.
\newblock PhD thesis, University of Calgary, 2005.

\end{thebibliography}

\renewcommand{\theequation}{A.\arabic{equation}}
\appendix
\section{Samples from UQ}
\begin{figure*}[h]
  \centering
   \includegraphics[width=0.9\linewidth]{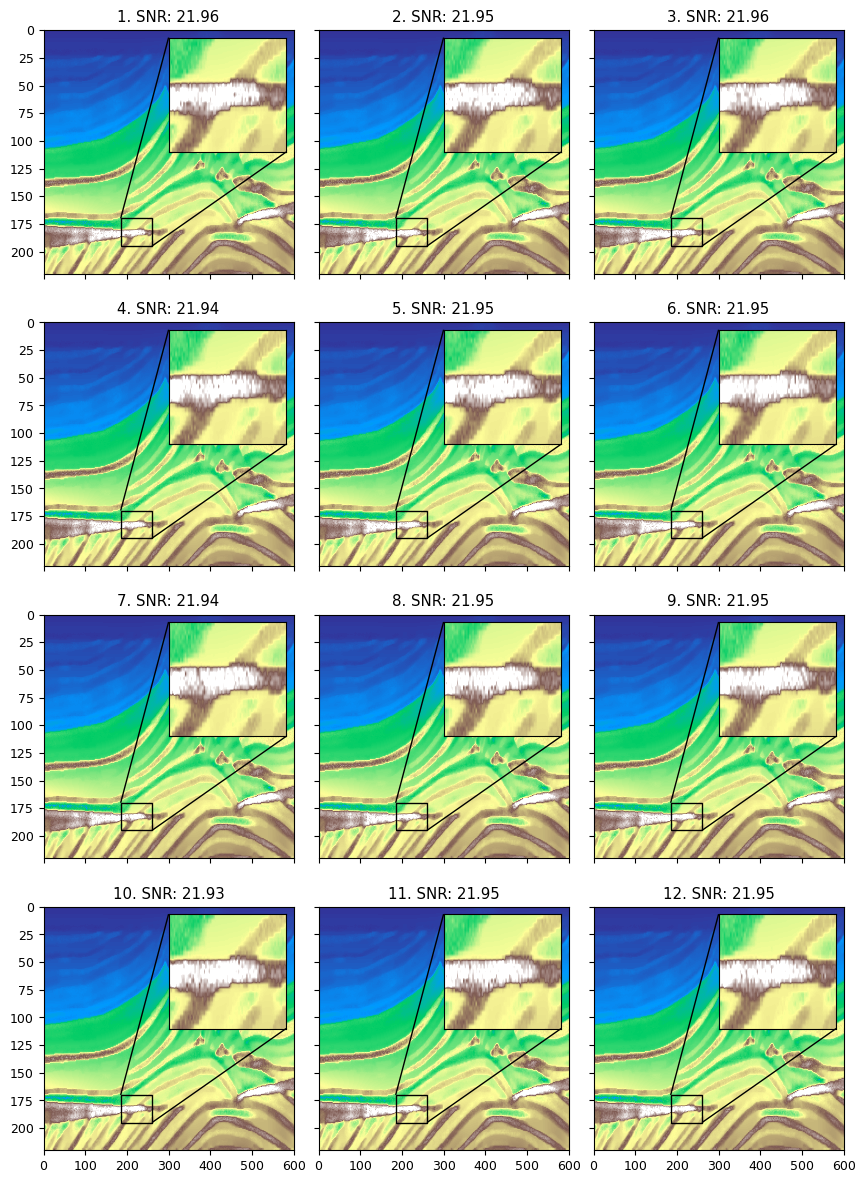}
   \caption{12 realizations of the Marmousi model obtained as part of the Monte Carlo Dropout uncertainty quantification process.}
   \label{fig:marm_uq_realizations}
\end{figure*}

Figure \ref{fig:marm_uq_realizations} displays 12 of the 100 realizations produced as part of the uncertainty quantification experiment conducted using IntraSeismic with Monte Carlo dropout (MCD) on the Marmousi model \cite{brougois_marmousi_1990}. Overall, the differences between the different models are not readily apparent. However, a detailed examination of each image reveals small variations at pixel resolution within the high-impedance structures, which are the areas of the largest uncertainty.

\section{Ablation study of IntraSeismic loss function components}
\begin{figure*}[h]
  \centering
   \includegraphics[width=\linewidth]{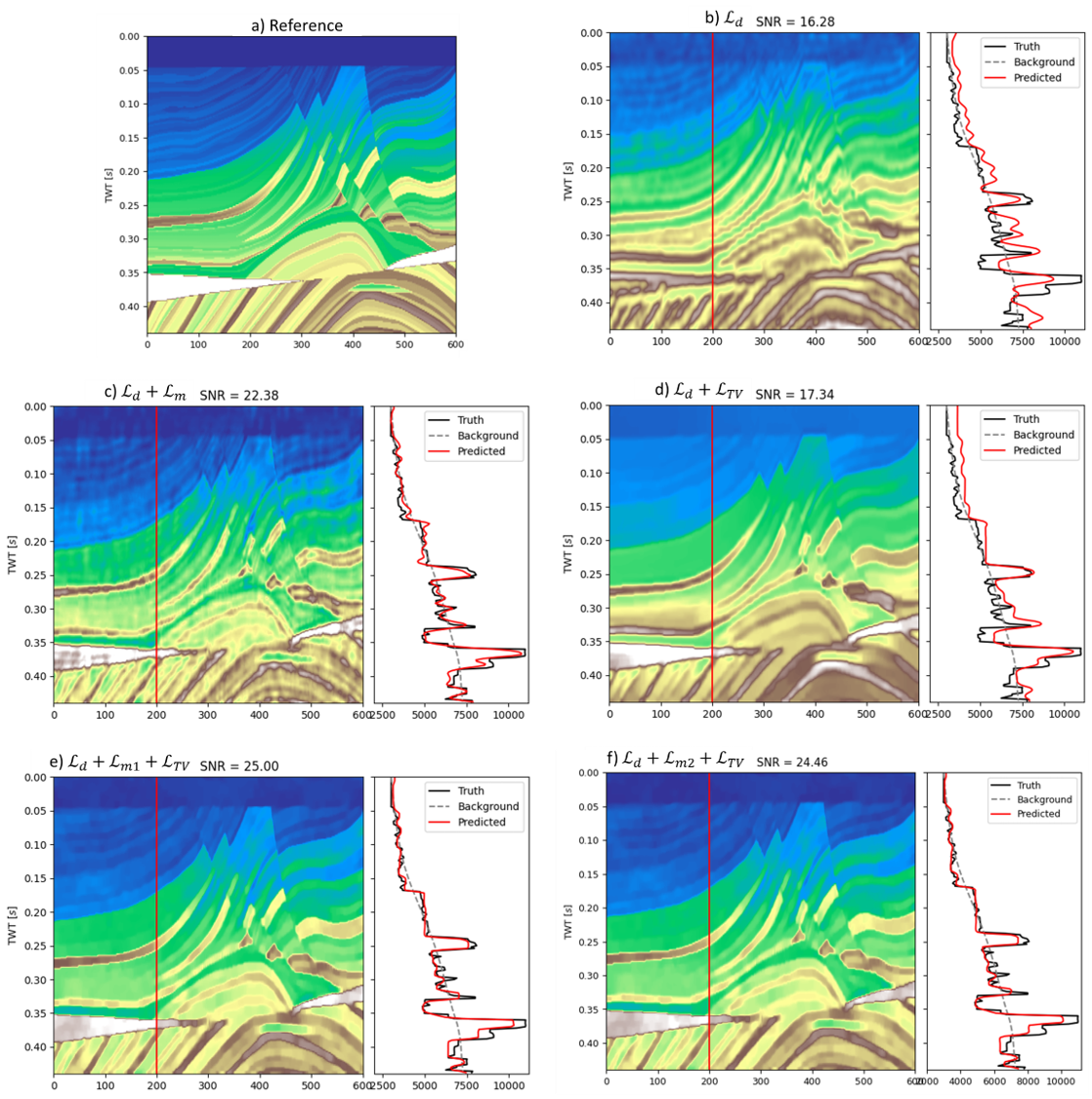}
   \caption{IntraSeismic results on Marmousi model inversion using different loss functions: a) Ground truth Marmousi model, b) data loss alone $\mathcal{L}_d$, c) data loss and L1 loss of the IntraSeismic MLP output $\mathcal{L}_d + \mathcal{L}_m$,  d) data loss and TV regularization of the final model output $\mathcal{L}_d + \mathcal{L}_{TV}$, e) full IntraSeismic loss $\mathcal{L}_d + \mathcal{L}_{m1} + \mathcal{L}_{TV}$ being $\mathcal{L}_{m}$ the L1 norm of the MLP output, and f) full IntraSeismic loss $\mathcal{L}_d + \mathcal{L}_{m2} + \mathcal{L}_{TV}$ being $\mathcal{L}_{m2}$ the L2 norm of the MLP output.}
   \label{fig:marm_abl}
\end{figure*}

A quantitative analysis has been conducted to investigate the impact that every loss term in the IntraSeismic loss function has on the final inverted acoustic impedance model. To summarize, the loss function is composed of three main terms as follows (for more details, see Section \ref{methodology}):

\begin{equation}
    \begin{aligned}
    \argmin_{\Theta} \frac{1}{2}\underbrace{\left\|\mathbf{G} \boldsymbol{m}_{\Theta} - \mathbf{d}\right\|_2^2}_{\text{\large $\mathcal{L}_d$}} + 
    \underbrace{\alpha\left\|\boldsymbol{m}_{\Theta}\right\|_{TV}}_{\text{\large $\mathcal{L}_{TV}$}} + 
    \underbrace{\beta\left\|\mathbf{F}_{\Theta}(\boldsymbol{x})\right\|_{1}}_{\text{\large $\mathcal{L}_m$}} \\
    \label{eqn:intraseismic_loss_}
    \end{aligned}
\end{equation}
The impact of each loss term and various combinations of loss terms on the final inverted model is shown in Figure \ref{fig:marm_abl}. In the first scenario (Figure \ref{fig:marm_abl}b), employing only the data loss term ($\mathcal{L}_d$), resulted in the lowest SNR of 16.28. The model reconstructed in this case was of lower resolution compared to the ground truth. In the second scenario (Figure \ref{fig:marm_abl}c), which combined the data loss with the MLP outputs L1 loss ($\mathcal{L}_d + \mathcal{L}_m$), an improvement in SNR was observed compared to using the data loss alone. However, the resulting reconstructed model appeared to contain a clear imprint of the noise in the data. The third scenario (Figure \ref{fig:marm_abl}d), involving the sum of the data loss and the TV loss on the final model output ($\mathcal{L}_d + \mathcal{L}_{TV}$), yielded a high-resolution model estimate. This estimate accentuated discontinuities and sharp contrasts, but as observed in the profile view, it uniformly deviated from the background model, thus not honoring the prior knowledge. In the fourth scenario (Figure \ref{fig:marm_abl}e), where all loss terms were combined ($\mathcal{L}_d + \mathcal{L}_m + \mathcal{L}_{TV}$), the reconstruction achieved the highest SNR among all methods tested. This was reflected in a high-resolution model estimate.

Additionally, we explored the possibility to use an $L_2$ norm with a dynamically varying weight ($\beta$) across iterations ($k$) to regularize the MLP output, as opposed to the previously described $L_1$ norm: $\mathcal{L}_m = \beta_k\left\|\mathbf{F}_{\Theta}(\boldsymbol{x})\right\|_{2}^2$. This regularization strategy aims to prioritize the fitting of the MLP output to the background model ($\boldsymbol{m}_\Theta = \boldsymbol{m}_{\text{back}}$) in the initial iterations. As the weight $\beta$ decreases over time, the network progressively adapts to capture and fit the high-resolution features of the model. The effectiveness of this approach is illustrated in Figure \ref{fig:marm_abl} f. These results closely mirror those achieved with the $L_1$ norm, as depicted in Figure \ref{fig:marm_abl} e, although with lower SNR.

\section{IntraSeismic for pre-stack seismic data}
Angle-dependent seismic data can be modeled using a generalized form of the convolutional model described in Section 2. Here, the data $d(t, \theta)$ can be represented as the convolution of a source function or wavelet $w(t)$ and the angle-dependent earth P-wave reflectivity series $r_{pp}(t, \theta)$:

\begin{equation}
    d(t, \theta)=w(t) * r_{pp}(t, \theta)
    \label{eqn:prestack}
\end{equation}

The Aki-Richards approximation, which builds on the foundational Zoeppritz equation \cite{aki2002, Zoeppritz1919}, expresses $r_{pp}$ as follows:
\begin{equation}
    r_{pp}(t, \theta)=\sum_{i=1}^3 c_i(t, \theta) \frac{d}{d t} \log \left(m_i(t)\right),
    \label{eqn:prestack_reflec}
\end{equation}
where the mixing coefficients $c_i$ are given by
\begin{equation}
    \begin{aligned}
    & c_1(t, \theta)=\frac{1}{2}\left(1+\tan ^2(\theta)\right) \\
    & c_2(t, \theta)=-4\left(\frac{\bar{V}_s(t)}{\bar{V}_p(t)}\right)^2 \sin ^2(\theta) \\
    & c_3(t, \theta)=\frac{1}{2} - 2\left(\frac{\bar{V}_s(t)}{\bar{V}_p(t)}\right)^2 \sin ^2(\theta)
    \end{aligned}
\end{equation}
and
\begin{equation}
    \mathbf{m}(t)=\left(V_p(t), V_s(t), \rho(t)\right),
\end{equation}
where $\theta$ represents the incidence angle of the seismic ray-path, $V_p$ is the P-wave velocity, $V_s$ the S-wave velocity, and $\rho$ is the density of the medium of interest. Note that equation \ref{eqn:zero_offset}, also known as the post-stack seismic modeling equation, pertains to the special scenario where $\theta = 0$, a condition termed as zero-offset data. Combining equations \ref{eqn:prestack} and \ref{eqn:prestack_reflec}, the pre-stack seismic inverse problem can be expressed as:

\begin{equation}
    \mathbf{m^*} = \argmin_{\mathbf{m}} \frac{1}{2}\left\|\mathbf{W C D m} - \mathbf{d}\right\|_2^2 + \mathcal{R}(\mathbf{m}),
    \label{eqn:prestack_inv}
\end{equation}
where $\textbf{C}$ is a matrix that contains the mixing coefficients.

\begin{figure*}[h]
  \centering
   \includegraphics[width=\linewidth]{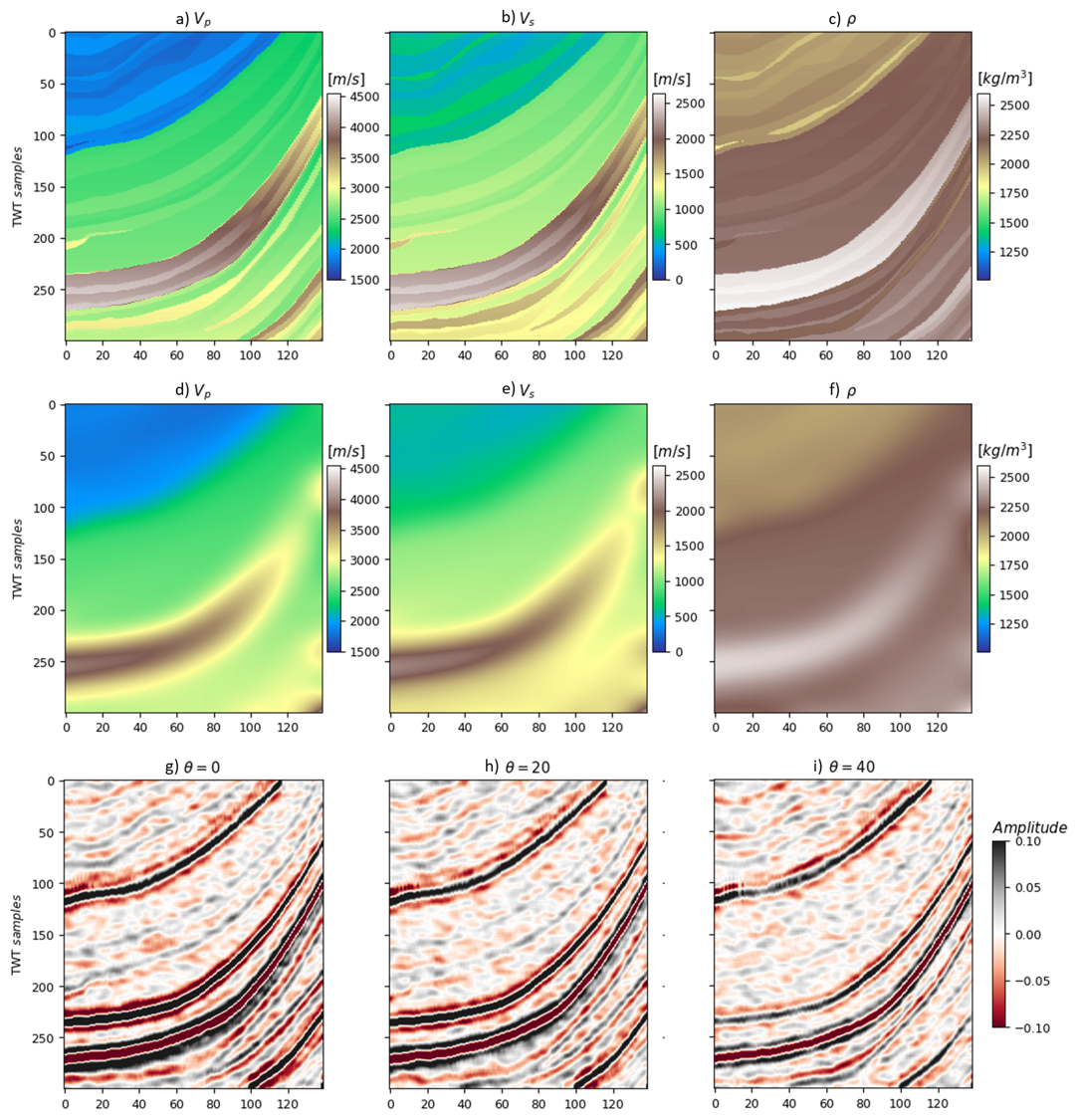}
   \caption{Pre-stack Marmousi dataset: a-c ground truth parameters $V_p, V_s$ and $\rho$, d-f background models, g-i partial stacks for angles $\theta=0$, $\theta=20$ and $\theta=40$.}
   \label{fig:prestack_data}
\end{figure*}

IntraSeismic can be seamlessly extended to directly invert pre-stack seismic data, simultaneously retrieving elastic properties such as $V_p$, $V_s$, and $\rho$. To showcase the applicability of IntraSeismic to pre-stack inversion, we use a portion of the synthetic elastic Marmousi model \cite{gary2006}, shown in Figures \ref{fig:prestack_data} a-c. Figures \ref{fig:prestack_data} d-f illustrate the background models used as starting guesses for the inversion, which are obtained by smoothing the corresponding ground truth models. Figures \ref{fig:prestack_data} g-i present three of the twenty-one computed common-angle gathers (i.e., pre-stack data at a fixed angle).

When using IntraSeismic to invert the three subsurface models simultaneously, as is the case for pre-stack seismic inversion, two approaches can be considered: 1) employing the standard IntraSeismic architecture to produce three network outputs, which are then processed by the physical modeling module (as depicted in Figure \ref{fig:arch}); 2) enhancing IntraSeismic with three networks (consisting of a hash table and MLP each), which individually output a different elastic property. Their outputs are subsequently concatenated and integrated into the physics modeling stage.

\begin{figure*}[h]
  \centering
   \includegraphics[width=\linewidth]{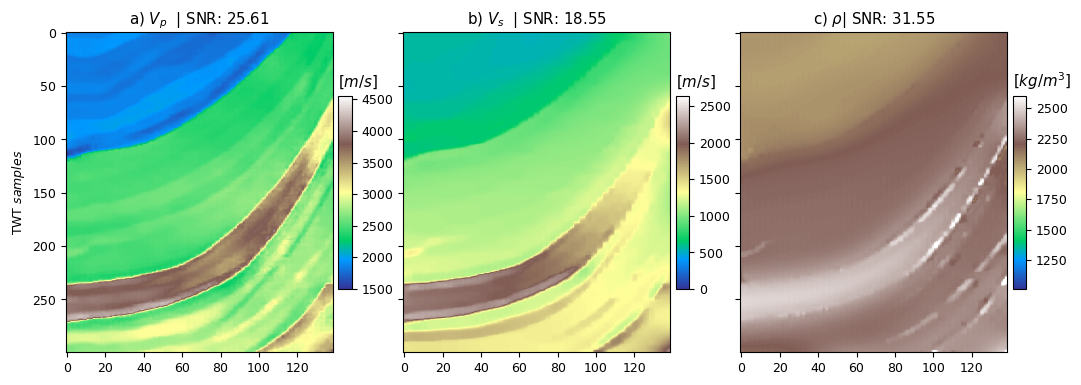}
   \caption{Pre-stack inversion results of a portion of the elastic Marmousi model using IntraSeismic: a) P-wave velocity $V_p$, b) S-wave velocity $V_s$, and c) density $\rho$.}
   \label{fig:prestack_inv1}
\end{figure*}

\begin{figure*}[h]
  \centering
   \includegraphics[width=\linewidth]{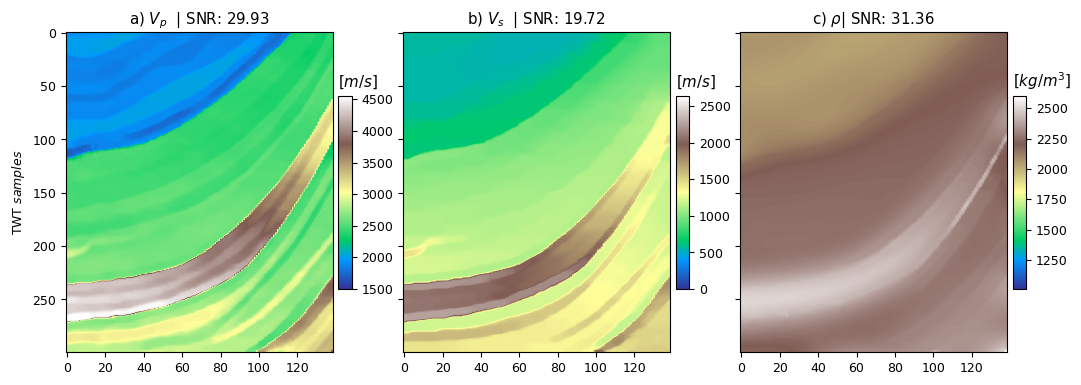}
   \caption{Pre-stack inversion results of a portion of the elastic Marmousi model using IntraSeismic with three separate networks: a) P-wave velocity $V_p$, b) S-wave velocity $V_s$, and c) density $\rho$.}
   \label{fig:prestack_inv2}
\end{figure*}

\begin{figure*}[h]
  \centering
   \includegraphics[width=\linewidth]{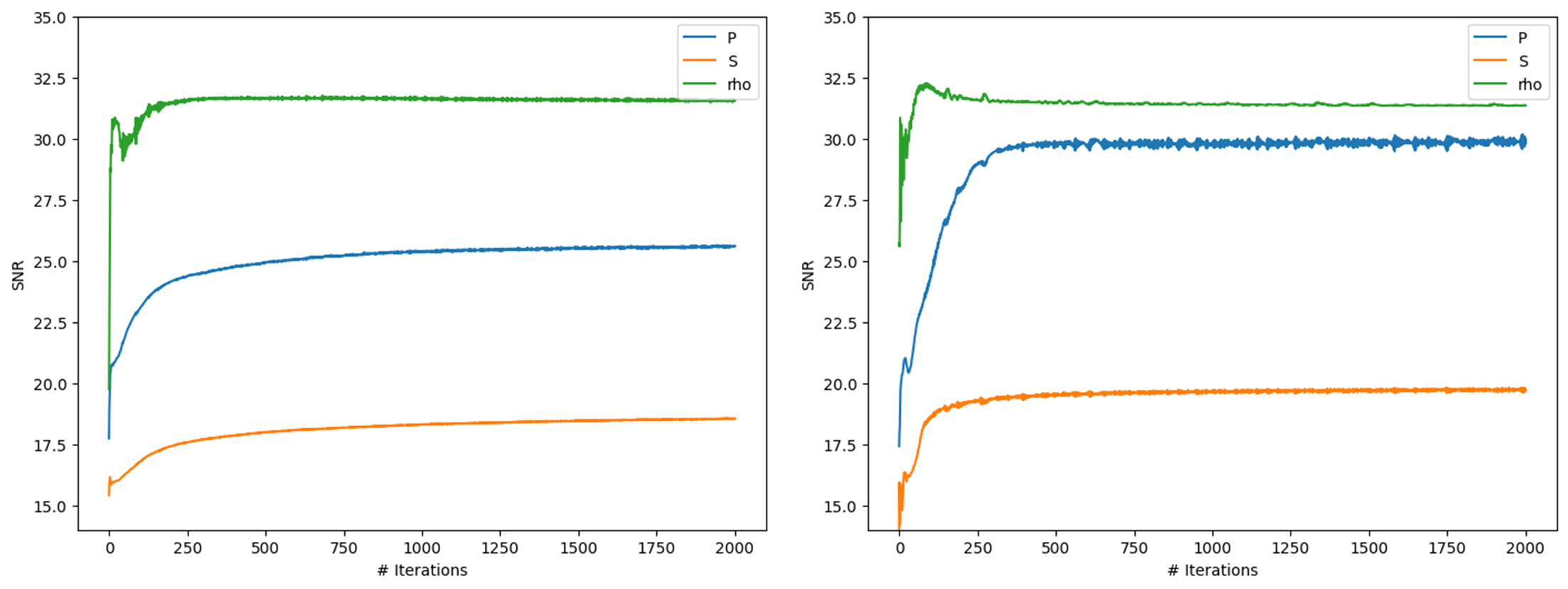}
   \caption{SNR evolution through iterations for each elastic parameter ($V_p, V_s, \rho$) using a) standard IntraSeismic, b) IntraSeismic with three networks.}
   \label{fig:marm_prestack_snr}
\end{figure*}

Figure \ref{fig:prestack_inv1} presents the outcomes of the first approach, showcasing a reasonable reconstruction of $V_p$ and $V_s$, although the latter displays a lower SNR. The estimation of $\rho$, on the other hand, although it appears notably noisier than the other two, achieves a higher SNR; this is attributed to the limited value range (2000 to 2500 $km/m^3$), resulting in a  naturally lower Total Variation (TV) term compared to other inverted properties. Conversely, Figure \ref{fig:prestack_inv2} depicts the results from the second approach, yielding a more effective reconstruction of $V_p$ and $V_s$. This is evident from their visual alignment with the ground truth models (Figures \ref{fig:prestack_data} a-c) and an improved SNR compared to the first approach. However, similar to the first method, the $\rho$ estimate is less accurate compared to $V_p$ and $V_s$, resembling a lower-resolution version of the ground truth. This issue is well known pre-stack seismic inversion and is not specific to IntraSeismic, owing to the limited sensitivity of pre-stack seismic data to density variations \cite{Downton}. Consequently, modifying the regularization weights for each parameter inevitably improves one model at the cost of others. Figure \ref{fig:marm_prestack_snr} illustrates the SNR evolution over iterations for each model in both approaches. Both approaches demonstrate swift convergence for the three models, with the SNR peaks occurring nearly simultaneously and exhibiting stable behavior after convergence.

\section{IntraSeismic sensitivity to network's initialization}
\begin{figure*}[h]
  \centering
   \includegraphics[width=\linewidth]{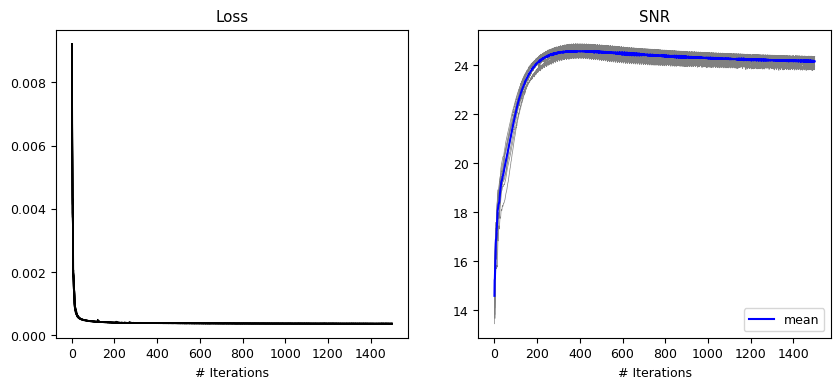}
   \caption{Results of 10 random initializations of IntraSeismic: a) loss function curves and b) SNR curves.}
   \label{fig:marm_init}
\end{figure*}

Finally, we evaluated the impact of the initialization of the parameters in the hash tables and MLP layers of IntraSeismic. We conducted ten separate runs for the Marmousi model, each with a different random seed. The resulting loss functions, displayed in Figure \ref{fig:marm_init}a, demonstrate a high degree of alignment across all runs, indicating a consistent behavior regardless of the initialization used. The evolution of the SNR for the different runs exhibits more variability (Figure \ref{fig:marm_init}b). Despite this variability, the final SNR for all runs consistently exceeded a value of 24dB, which is a notable improvement over standard methods, as detailed in Section \ref{experiments}.

\end{document}